\documentclass[aps, prd, twocolumn, superscriptaddress]{revtex4-1}
\usepackage{graphicx}
\usepackage{amssymb}
\usepackage{amsmath}
\usepackage{mathtools}
\usepackage{lmodern}
\usepackage{booktabs}
\usepackage{bbold}
\usepackage{bm}
\usepackage{bbm}
\usepackage[colorlinks=True,linkcolor=blue,anchorcolor=blue,citecolor=blue,CJKbookmarks=true]{hyperref}

\usepackage{ulem}

\begin{document}
\title{Angular momentum transfer in multiphoton pair production}

\author{R. Z. Jiang}
\affiliation{State Key Laboratory for Tunnel Engineering, China University of Mining and Technology, Beijing 100083, China}


\author{Z. L. Li}
\email{Contact author: zlli@cumtb.edu.cn}
\affiliation{State Key Laboratory for Tunnel Engineering, China University of Mining and Technology, Beijing 100083, China}
\affiliation{School of Science, China University of Mining and Technology, Beijing 100083, China}


\author{Y. J. Li}
\email{Contact author: lyj@aphy.iphy.ac.cn}
\affiliation{State Key Laboratory for Tunnel Engineering, China University of Mining and Technology, Beijing 100083, China}
\affiliation{School of Science, China University of Mining and Technology, Beijing 100083, China}

\date{\today}

\begin{abstract}
We propose a method to accurately calculate the momentum distributions and the phase distributions of the probability amplitude for both boson and fermion pair production in a spatially homogeneous and time-dependent electric field.
Applying this method to multiphoton pair production in a circularly polarized electric field rotating around the $z$-axis, we clarify that the topological charge appearing in the phase distribution of the probability amplitude for pair production reflects the orbital angular momentum (OAM) of the produced pairs rather than that of individual particles.
On this basis, we demonstrate that, within the semiclassical framework, the $z$-component of the total angular momentum of the field and the particles is conserved, whereas the conservation of total angular momentum cannot be verified.
The results also reveal that the pair production is also constrained by $C$-parity conservation, and that pairs with smaller OAM are produced more favorably.
These findings provide deeper insight into angular momentum transfer in pair production.
\end{abstract}

\maketitle

\section{INTRODUCTION}
\label{sec:one}
In $1928$, the formulation of the Dirac equation \cite{Dirac1928} transformed the spin angular momentum of electrons from a hypothesis into a natural consequence of relativistic quantum mechanics, while another significant contribution of this equation was its first prediction of the existence of the positron.
Sauter solved the Dirac equation in $1931$ for the presence of a strong static electric field \cite{Sauter1931}, discovering the phenomenon of crossing between positive and negative energy continua.
This indicates that in the Dirac sea model, negative energy electrons can tunnel into positive energy states and produce electron-positron pairs.
In $1951$, Schwinger calculated the production rate of electron-positron pairs in a constant electric field of strength $E$ as $\mathcal{P}\sim\exp(-\pi E_{\mathrm{cr}}/E)$ using the proper time method \cite{Schwinger1951}, where $E_{\mathrm{cr}}=m^2/e\sim 1.3\times 10^{16}\mathrm{V}/\mathrm{cm}$ is the critical electric field strength, $m$ and $e$ represent the magnitude of mass and charge for an electron, respectively.
In this paper, natural units $\hbar=c=1$ are used.
Due to these contributions, the phenomenon of the quantum vacuum becoming unstable and decaying into electron-positron pairs in a strong background field is also known as the Sauter-Schwinger effect.
In 1970, Brezin and Itzykson employed the Wentzel-Kramers-Brillouin (WKB) approximation to investigate the creation of particle-antiparticle pairs in a spatially homogeneous and time-dependent sinusoidal electric field \cite{Brezin1970P}, and pointed out that the Keldysh adiabatic parameter $\gamma_{\mathrm{K}}=m\Omega/(eE)$ \cite{Keldysh1965i} can be used to distinguish between two limiting scenarios for pair production.
Here $\Omega$ denotes the frequency of the electric field.
The limits where $\gamma_{\mathrm{K}}\ll 1$ and $\gamma_{\mathrm{K}}\gg 1$ are referred to as the Schwinger tunneling mechanism and the multiphoton absorption process, respectively.

So far, the Sauter-Schwinger effect has not received direct experimental verification, because the laser field intensity $I_{\mathrm{cr}}\sim 4.3\times 10^{29}\mathrm{W}/\mathrm{cm}^2$  corresponding to the Schwinger critical electric field $E_{\mathrm{cr}}$ far exceeds current laboratory capabilities.
Fortunately, with the rapid development of laser technology, high-power lasers are expected to achieve intensities of $10^{26}\mathrm{W}/\mathrm{cm}^2$ in the future \cite{ELI,XCELS}.
The Sauter-Schwinger effect has become one of the research focuses in quantum electrodynamics (QED).
Current research is primarily focused along two main directions.
One aims to enhance the pair yield under subcritical electric fields below a threshold of $0.1E_{\mathrm{cr}}$.
To this end, methods such as the dynamically assisted Sauter-Schwinger mechanism \cite{Schutzhold2008D,Li2014E,Schneider2016P,Otto2014L,Aleksandrov2018D,Edwards2025R}, chirped electric fields \cite{Abdukerim2017E,Olugh2018P,Gong2019E,Chen2024A}, and optimized electric field pulses \cite{Kohlfurst2012O,Hebenstreit2014O} have been proposed.
The other direction focuses on investigating the physical phenomena in vacuum pair production.
This not only enables a deeper theoretical understanding of the mechanism of the Sauter-Schwinger effect but also provides a crucial foundation for interpreting experimental results.

Studying the Sauter-Schwinger effect through the momentum distribution function is a highly effective approach that has revealed many important signatures in pair production, such as effective mass features \cite{Kohlfurst2013E}, momentum spirals \cite{Li2017M,Li2019B,Hu2023M}, time-domain multi-slit interference effects \cite{Akkermans2011R,Li2014xga}, nodal structures \cite{Li2015N,Fan2024vqi}, and so on.
In fact, the discovery of these important features generally does not require consideration of the spin angular momentum of particles. The spin of particles is treated as degenerate.
In recent years, the physical phenomena arising from spin effects in vacuum pair production have become a topic of growing interest.
Researchers have found that even when using the standing wave approximation for the field and neglecting its magnetic component, observable spin asymmetry and a variety of other spin-related phenomena still emerge \cite{Wollert2015kra,Blinne2015zpa,Li2019B,Kohlfurst2018S,Hu2024nyp,
Aleksandrov2024cqh,Majczak2024hmt,Amat2024nvg,Chen2025xib}.
In Refs. \cite{Majczak2024hmt,Amat2024nvg}, the spin-resolved momentum distribution for produced pairs was obtained by using the scattering matrix approach and the two-level model, respectively.
Compared to the Dirac-Heisenberg-Wigner (DHW) formalism \cite{Hu2024nyp,Chen2025xib}, the advantage of these two methods lies not only in considering the spins of electrons and positrons simultaneously but also in enabling the study of the phase distribution of the probability amplitude for pair production, which contains rich physical information.
Using the two-level model, the phase distribution of the probability amplitude for multiphoton scalar pair production is computed in Ref. \cite{Fan2024nsl}.
The result shows that the topological charge in the phase distribution reflects the OAM of produced particles, and angular momentum is conserved in multiphoton scalar pair production.
However, in the scattering matrix approach, the probability amplitude for pair production contains the dynamical phase factor that must be determined empirically.
In the two-level model, to derive the two-level formula for fermion pair production, the inner product terms between the time derivatives of the instantaneous eigenvectors and the instantaneous eigenvectors themselves are neglected.
Both methods have their shortcomings.

In this paper, we presents an accurate approach to calculate the momentum distributions and the phase distributions of the probability amplitude for boson and fermion pair production.
In our approach, the dynamical phase factor is explicitly defined, and the phase distributions are independent of the switching on and off times of the electric field, offering excellent observability.
Employing this approach, we clarify the relation between the topological charge and orbital angular momentum, and examine the angular momentum transfer from field to particles for both boson and fermion pair production.
The study will further deepen the understanding of angular momentum transfer in multiphoton pair production.

The structure of this paper is as follows:
In Sec. \ref{sec:two}, we introduce the method for calculating the probability amplitude for boson and fermion pair production under a spatially homogeneous and time-dependent electric field.
In Sec. \ref{sec:three}, we numerically investigate the momentum distributions and the phase distributions of probability amplitude for boson and fermion pair production in a circularly polarized electric field rotating around $z$-axis.
The summary is given in Sec. \ref{sec:four}.
Appendix \ref{appa} shows the Dirac-Heisenberg-Wigner formalism for calculating the spin-resolved momentum distribution.

\section{THEORETICAL FORMULA}
\label{sec:two}
In this section, we will propose a method for calculating the probability amplitudes and corresponding phases for boson and fermion pair production from vacuum under a spatially homogeneous and time-dependent electric field.

\subsection{Boson pair production from vacuum}\label{subsec:A}
For boson pair production, our starting point is the Klein-Gordon equation in the presence of an electromagnetic field
\begin{eqnarray}\label{eq:KLEIN}
\left( D^{\mu}D_{\mu}+m^2 \right) \Phi \left( \mathbf{x}, t \right) =0,
\end{eqnarray}
where $D_{\mu}=\partial_{\mu}+ieA_{\mu}(\mathbf{x}, t)$ is the covariant derivative, $A_{\mu}(\mathbf{x}, t)$ is the four-vector potential of the electromagnetic field, and $e$ and $m$ represent the charge and mass of the particle, respectively.
For computational convenience, the Klein-Gordon equation can be equivalently transformed into a Schr\"{o}dinger-type equation within the Feshbach-Villars representation \cite{Feshbach1958}:
\begin{eqnarray}\label{eq:KLEINFV}
i\partial _t\phi \left( \mathbf{x}, t \right) =\hat{H}\left( \mathbf{x}, t \right) \phi \left( \mathbf{x}, t \right),
\end{eqnarray}
where
\begin{equation}\label{eq:FVF}
\phi \left( \mathbf{x}, t \right) =\frac{1}{2}\left( \begin{array}{c}
\Phi +\frac{i}{m}\frac{\partial \Phi}{\partial t}-\frac{eA^0}{m}\Phi\\
\Phi -\frac{i}{m}\frac{\partial \Phi}{\partial t}+\frac{eA^0}{m}\Phi\\
\end{array} \right),
\end{equation}
\begin{eqnarray}\label{eq:FVH}
\hat{H}=\frac{\left( \hat{\mathbf{q}}-e\mathbf{A} \right) ^2}{2m}\left( \sigma _3+i\sigma _2 \right) +m\sigma _3+eA^0\mathbbm{1}_2
\end{eqnarray}
is the Feshbach-Villars Hamiltonian, $\hat{\mathbf{q}}=-i\nabla$ is the canonical momentum operator, $\mathbbm{1}_2$ is a $2\times2$ identity matrix, and $\boldsymbol{\sigma}=(\sigma_1, \sigma_2, \sigma_3)$ is the Pauli matrix.

In the following, we consider a spatially homogeneous and time-dependent electric field $\mathbf{E}(t)$. The corresponding four-vector potential under temporal gauge is $A_{\mu}(\mathbf{x}, t)=(0, -\mathbf{A}(t))$ and $\mathbf{E}(t)=-\dot{\mathbf{A}}(t)$.
Here the dot above the letter denotes the derivative with respect to time.
For this electric field, the canonical momentum $\mathbf{q}$ is a good quantum number, then the Fourier expansion of $\phi(\mathbf{x}, t)$ is
\begin{eqnarray}\label{eq:FEOK}
\phi \left( \mathbf{x}, t \right) =\int{\left[ dq \right]}e^{i\mathbf{q}\cdot \mathbf{x}}\varphi _{\mathbf{q}}\left( t \right),
\end{eqnarray}
where $\varphi_{\mathbf{q}}(t)$ denotes the Fourier modulus, and $[dq]=d^3q/(2\pi)^3$.
Inserting the above equation into Eq. (\ref{eq:KLEINFV}), we obtain
\begin{subequations}\label{eq:FVFFMS}
\begin{align}
& i\partial _t\varphi _{\mathbf{q}}\left( t \right) =H_{\mathbf{p}} \varphi _{\mathbf{q}}\left( t \right) , \label{eq:FVFFMS_a} \\
& H_{\mathbf{p}}=\frac{\mathbf{p}^2}{2m}\left( \sigma _3+i\sigma _2 \right) +m\sigma _3, \label{eq:FVFFMS_b}
\end{align}
\end{subequations}
where $H_{\mathbf{p}}$ is the Hamiltonian in momentum space and $\mathbf{p}=\mathbf{q}-e\mathbf{A}(t)$ represents the kinetic momentum. The orthonormal instantaneous eigenstates of $H_{\mathbf{p}}$ are
\begin{eqnarray}\label{eq:PANES}
\begin{array}{l}
\widetilde{\varphi}_{\mathbf{q},+}(t)=\frac{1}{\sqrt{4m\omega _{\mathbf{q}}\left( t \right) }}\left( \begin{array}{c}
	\omega _{\mathbf{q}}\left( t \right) +m\\
	-\omega _{\mathbf{q}}\left( t \right) +m\\
\end{array} \right)e^{-i\vartheta \left( t \right)}  ,
\\
\\
\widetilde{\varphi}_{-\mathbf{q},-}(t)=\frac{1}{\sqrt{4m\omega _{\mathbf{q}}\left( t \right) }}\left( \begin{array}{c}
	-\omega _{\mathbf{q}}\left( t \right) +m\\
	\omega _{\mathbf{q}}\left( t \right) +m\\
\end{array} \right)e^{i\vartheta \left( t \right)} ,
\end{array}
\end{eqnarray}
and satisfy $H_{\mathbf{p}}\widetilde{\varphi}_{\mathbf{q},+}(t)= \omega_{\mathbf{q}}(t)\widetilde{\varphi}_{\mathbf{q},+}(t)$ and $H_{\mathbf{p}}\widetilde{\varphi}_{-\mathbf{q},-}(t)= -\omega_{\mathbf{q}}(t)\widetilde{\varphi}_{-\mathbf{q},-}(t)$, where $\omega _{\mathbf{q}}\left( t \right)=\sqrt{[\mathbf{q}-e\mathbf{A}(t)]^2+m^2}$ denotes the energy of the particle and
\begin{align}\label{eq:DPAIP}
\vartheta \left( t \right) =\int_{t_{\mathrm{in}}}^t{\omega _{\mathbf{q}}\left( \tau \right) d\tau}+\omega _{\mathbf{q}}\left( t_{\mathrm{in}} \right) t_{\mathrm{in}}
\end{align}
is the instantaneous phase which includes the dynamical phase and the initial phase.


Since the two independent solutions ($\varphi _{\mathbf{q},+}$ and $\varphi _{-\mathbf{q},-}$) of Eq. (\ref{eq:FVFFMS_a}) and the instantaneous eigenstates (\ref{eq:PANES}) are both complete bases in pseudospinor space, the Feshbach-Villars field operator $\hat{\phi} \left( \mathbf{x},t \right)$ can be expanded separately using them as
\begin{eqnarray}\label{eq:EPBAA}
\hat{\phi} \left( \mathbf{x}, t \right) =&&\int{\left[ dq \right]}e^{i\mathbf{q}\cdot \mathbf{x}}\left[ \hat{a}_{\mathbf{q},+}\varphi _{\mathbf{q},+}\left( t \right) +\hat{a}_{-\mathbf{q},-}^{\dagger}\varphi_{-\mathbf{q},-}\left( t \right) \right],\nonumber \\
=&&\int{\left[ dq \right]}e^{i\mathbf{q}\cdot \mathbf{x}}\big[\hat{a}_{\mathbf{q},+}\left( t \right) \widetilde{\varphi}_{\mathbf{q},+}\left( t \right) \\
 &&\hspace{1.8cm}+ \hat{a}_{-\mathbf{q},-}^{\dagger}\left( t \right) \widetilde{\varphi} _{-\mathbf{q},-}\left( t \right) \big], \nonumber
\end{eqnarray}
where
$\hat{a}_{\mathbf{q},+}$ and $\hat{a}_{-\mathbf{q},-}^{\dagger}$ ($\hat{a}_{\mathbf{q},+}(t)$ and $\hat{a}_{-\mathbf{q},-}^{\dagger}(t)$) are the time-independent (time-dependent) particle annihilation operator and antiparticle creation operator, respectively. The time-independent operators satisfy $\hat{a}_{\mathbf{q},+}|\mathrm{vac}\rangle=0$ and $\langle \mathrm{vac}|\hat{a}_{-\mathbf{q},-}^{\dagger}=0$, where $|\mathrm{vac}\rangle$ represents the vacuum state.

The probability density of the produced bosons is defined as
\begin{align}\label{eq:EPBATAT}
\rho_{\mathrm{B}}(\mathbf{x}, t)=\langle \mathrm{vac}|\hat{\phi} _{+}^{\dagger}\left( \mathbf{x}, t \right) \sigma _3\hat{\phi} _+\left( \mathbf{x}, t \right) |\mathrm{vac}\rangle,
\end{align}
where $\hat{\phi}_+(\mathbf{x}, t)=\int{[dq]}e^{i\mathbf{q}\cdot \mathbf{x}}\hat{a}_{\mathbf{q},+}(t)\widetilde{\varphi}_{\mathbf{q},+}(t)$ represents the positive energy part of the field operator.
According to Eqs. (\ref{eq:EPBAA}) and (\ref{eq:EPBATAT}), we can obtain the particle number density
\begin{eqnarray}\label{eq:NTDOB}
n_{\mathrm{B}}\left( t \right) =
\int d^3x \rho_\mathrm{B}(\mathbf{x},t)=
\int{\left[ dq \right] f_{\mathrm{B}}\left( \mathbf{q},t \right)},
\end{eqnarray}
where
\begin{equation}\label{eqn:MD-B}
f_{\mathrm{B}}(\mathbf{q},t)=|c_{\mathrm{B}}(\mathbf{q},t)|^2
\end{equation}
is the single-particle momentum distribution function of produced bosons, and
\begin{equation}\label{eqn:prob-B}
c_{\mathrm{B}}(\mathbf{q},t)=\widetilde{\varphi} _{\mathbf{q},+}^{\dagger}(t)\sigma _3\varphi _{-\mathbf{q},-}(t)
\end{equation}
is the probability amplitude for boson pair production.
$\varphi _{-\mathbf{q},-}(t)$ can be obtained by evolving $\widetilde{\varphi} _{-\mathbf{q},-}(t_{\mathrm{in}})$ according to Eq. (\ref{eq:FVFFMS_a}).
Note that the created particles should be understood as quasiparticles in the presence of an external field.
To obtain the number density of real particles, the external field should be asymptotically approaches $0$.

\subsection{Fermion pair production from vacuum}\label{subsec:B}
To obtain the spin-resolved probability amplitudes and momentum distribution functions of electron-positron pairs produced from vacuum in a strong background field, we start from the Dirac equation in the presence of an electromagnetic field
\begin{eqnarray}\label{eq:DIRACEQ}
\left[ i\gamma ^{\mu}\partial _{\mu}-e\gamma ^{\mu}A_{\mu}\left( \mathbf{x}, t \right) -m \right] \Psi \left( \mathbf{x}, t \right)=0,
\end{eqnarray}
where $\gamma ^{\mu}=(\gamma ^0, \boldsymbol{\gamma})$ is the Dirac matrix, and in the Dirac representation
\begin{eqnarray}\label{eq:DIRACM}
\gamma ^0=\left( \begin{matrix}
	\mathbbm{1}_2&		0\\
	0&		-\mathbbm{1}_2\\
\end{matrix} \right) ,    \boldsymbol{\gamma }=\left( \begin{matrix}
	0&		\boldsymbol{\sigma }\\
	-\boldsymbol{\sigma }&		0\\
\end{matrix} \right).
\end{eqnarray}

For a spatially homogeneous and time-dependent electric field, similar to the boson case, we take the following Fourier transform of the Dirac field
\begin{eqnarray}\label{eq:FTOD}
\Psi \left( \mathbf{x}, t \right) =\int{\left[ dq \right]}e^{i\mathbf{q}\cdot \mathbf{x}}\psi _{\mathbf{q}}\left( t \right),
\end{eqnarray}
where $\psi_{\mathbf{q}}(t)$ is the Fourier modulus.
Inserting the above equation into Eq. (\ref{eq:DIRACEQ}), we have
\begin{subequations}\label{eq:DIRACFM}
\begin{align}
& i\partial _t\psi _{\mathbf{q}}\left( t \right) =\mathcal{H} _{\mathbf{p}} \psi _{\mathbf{q}}\left( t \right) , \label{eq:DIRACFM_a} \\
& \mathcal{H} _{\mathbf{p}}=\gamma ^0\boldsymbol{\gamma }\cdot \mathbf{p}+\gamma ^0m, \label{eq:DIRACFM_b}
\end{align}
\end{subequations}
where $\mathcal{H}_{\mathbf{p}}$ is the time-dependent Hamiltonian in momentum space. The orthonormal instantaneous eigenstates of  $\mathcal{H}_{\mathbf{p}}$ are
\begin{align}
\label{eq:FFHOPNE}
\begin{aligned}
&\widetilde{u}_{\mathbf{q},s}(t) = \sqrt{\frac{\omega_{\mathbf{q}}(t) + m}{2\omega_{\mathbf{q}}(t)}}
\begin{pmatrix}
    \chi_s \\[1.2ex]
    \dfrac{\boldsymbol{\sigma} \cdot \mathbf{p}}{\omega_{\mathbf{q}}(t) + m} \chi_s
\end{pmatrix}e^{-i\vartheta \left( t \right)}, \\[2.5ex]
&\widetilde{v}_{-\mathbf{q},s}(t) = \sqrt{\frac{\omega_{\mathbf{q}}(t) + m}{2\omega_{\mathbf{q}}(t)}}
\begin{pmatrix}
    -\dfrac{\boldsymbol{\sigma} \cdot \mathbf{p}}{\omega_{\mathbf{q}}(t) + m}  \chi_s \\[1.2ex]
     \chi_s
\end{pmatrix}e^{i\vartheta \left( t \right)},
\end{aligned}
\end{align}
and satisfy $\mathcal{H} _{\mathbf{p}}\widetilde{u}_{\mathbf{q},s}(t)=\omega _{\mathbf{q}}(t)\widetilde{u}_{\mathbf{q},s}(t)$
 and
$\mathcal{H} _{\mathbf{p}}\widetilde{v}_{-\mathbf{q},s}(t)=-\omega _{\mathbf{q}}(t)\widetilde{v}_{-\mathbf{q},s}(t)$, where $s$ denotes spin up ($s=1$) and spin down ($s=2$) along the $z$ direction, respectively,
$\chi_1=(1,0)^{\intercal}$ and $\chi_2=(0,1)^{\intercal}$.


Similar to the case of bosons, the Dirac field operator can be expanded by the independent solutions ($u_{\mathbf{q},s}$ and $v _{-\mathbf{q},s}$) of Eq. (\ref{eq:DIRACFM_a}) as well as the instantaneous eigenstates (\ref{eq:FFHOPNE}):
\begin{eqnarray}\label{eq:EDBBD}
\hat{\Psi} \left( \mathbf{x}, t \right) =&& \int{\left[ dq \right]} e^{i\mathbf{q}\cdot \mathbf{x}} \sum_{s=1}^2{\left[ \hat{b}_{\mathbf{q},s} u_{\mathbf{q},s}\left( t \right) + \hat{d}_{-\mathbf{q},s}^{\dagger} v_{-\mathbf{q},s}\left( t \right) \right]} \nonumber \\
=&& \int{\left[ dq \right]} e^{i\mathbf{q}\cdot \mathbf{x}} \sum_{s=1}^2{\Big[ \hat{b}_{\mathbf{q},s}\left( t \right) \widetilde{u}_{\mathbf{q},s}(t)}  \\
&& \hspace{2.6cm}+ \hat{d}_{-\mathbf{q},s}^{\dagger}\left( t \right)\widetilde{v}_{-\mathbf{q},s}(t) \Big], \nonumber
\end{eqnarray}
where $\hat{b}_{\mathbf{q},s}$ and $\hat{d}_{-\mathbf{q},s}^{\dagger}$ ($\hat{b}_{\mathbf{q},s}(t)$ and $\hat{d}_{-\mathbf{q},s}^{\dagger}(t)$) are the time-independent (time-dependent) electron annihilation and positron creation operators, respectively.
The time-independent operators satisfy $\hat{b}_{\mathbf{q},s}|\mathrm{vac}\rangle=0$ and $\langle \mathrm{vac}|\hat{d}_{-\mathbf{q},s}^{\dagger}=0$, where $|\mathrm{vac}\rangle$ is the corresponding vacuum state.

The probability density of electrons produced from the vacuum is defined as
\begin{eqnarray}\label{eq:DOENDFV}
\rho_{e^-}\left( \mathbf{x},t \right)=\langle \mathrm{vac}|\hat{\Psi} _{e^-}^{\dagger}\left( \mathbf{x}, t \right) \hat{\Psi}_{e^-}\left( \mathbf{x}, t \right) |\mathrm{vac}\rangle,
\end{eqnarray}
where $\hat{\Psi} _{e^-}\left( \mathbf{x},t \right)=\int{\left[ dq \right]} e^{i\mathbf{q}\cdot \mathbf{x}} \sum_{s=1}^2{ \hat{b}_{\mathbf{q},s}\left( t \right) \widetilde{u}_{\mathbf{q},s}(t)}$ represents the positive energy part of the field operator.
According to Eqs. (\ref{eq:EDBBD}) and (\ref{eq:DOENDFV}), we can ultimately obtain the number density of created electrons
\begin{equation}\label{eq:SRMDOF}
n_{e^-}\left( t \right) =\!\int d^3x \rho_{e^-}\left( \mathbf{x},t \right)
=\!\int{\left[ dq \right]}\sum_{s,s^{\prime}=1}^2{f_{s,s^{\prime}}\left( \mathbf{q},t \right)},
\end{equation}
where
\begin{equation}\label{eqn:MD-e}
f_{s,s^{\prime}}\left( \mathbf{q},t \right) =\left| c_{s,s^{\prime}}\left( \mathbf{q},t \right) \right|^2
\end{equation}
is the spin-resolved momentum distribution function of created electrons,
\begin{equation}\label{eqn:prob-e}
c_{s,s^{\prime}}\left( \mathbf{q},t \right) =\widetilde{u}_{\mathbf{q},s}^{\dagger}(t)v_{-\mathbf{q},s^{\prime}}\left( t \right)
\end{equation}
is the probability amplitude for electron-positron pair production.  
Note that $v _{-\mathbf{q},s^{\prime}}(t)$ can be obtained by evolving $\widetilde{v} _{-\mathbf{q},s^{\prime}}(t_{\mathrm{in}})$ according to Eq. (\ref{eq:DIRACFM_a}).

\begin{figure*}[!ht]
\centering
\includegraphics[width=0.66\textwidth]{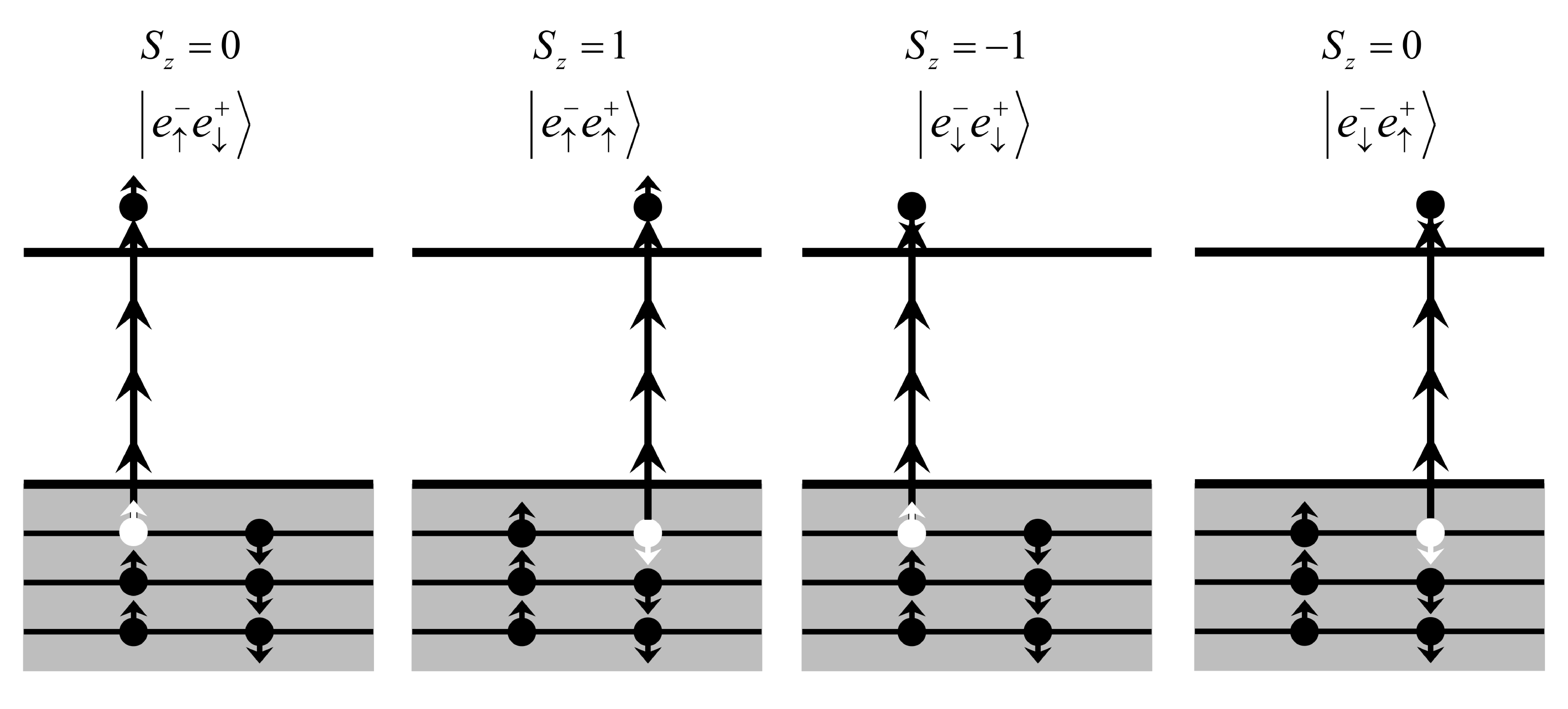}
\caption{Schematic diagram of the four possible transition modes for a negative-energy electron to the positive-energy continuum under an external field.
For the four spin configurations, the $z$-components of the spin angular momentum of the produced electron-positron pairs are $0$, $1$, $-1$, and $0$, respectively.
\label{fig:fouramp}}
\end{figure*}

Now we further explain the physical significance of the probability amplitude using the hole theory.
In this theory, according to the Pauli exclusion principle, each energy level in the Dirac sea is filled by a pair of negative-energy electrons with opposite spin projections, resulting in a net spin of $0$.
A spin-up (spin-down) negative-energy electron may transition to a positive-energy continuum under the influence of an external field, leaving behind a hole in the Dirac sea that is interpreted as a spin-down (spin-up) positron.
This is the picture of electron-positron pair production in the hole theory.
However, it should be noted that although a negative-energy electron has a definite spin, the spin of the produced positive-energy electron after its transition, under the influence of an external electric field, is still not fixed and may be either up or down.
Therefore, there are four possible transition modes for negative-energy electrons, as shown in Fig. \ref{fig:fouramp}.
Furthermore, the projection (\ref{eqn:prob-e}) of the evolved negative-energy state with spin label $s'$ onto the instantaneous positive-energy state with spin label $s$ represents the probability amplitude for producing an electron with spin label $s$ and a positron with spin label $\overline{s'}$, where $\overline{s'}$ is the opposite of $s'$.
Specifically, when a spin-up negative-energy electron ($s'=1$) transitions to a positive-energy state, a spin-down hole (positron, $\overline{s'}=2$) will be produced in the Dirac sea.
If the spin-up negative-energy electron transitions to a spin-up positive-energy electron ($s=1$), then an electron-positron pair with a total spin projection of $S_z=0$ is produced from vacuum, see the first panel in Fig. \ref{fig:fouramp}.
The probability amplitude for this process can be described by $c_{1,1}$.
If the spin-up negative-energy electron transitions to a spin-down positive-energy electron ($s=2$), then an electron-positron pair with a total spin projection of $S_z=-1$ is produced from vacuum, see the third panel in Fig. \ref{fig:fouramp}.
The probability amplitude for this process can be described by $c_{2,1}$.
Similarly, when a spin-down negative-energy electron transitions to the positive-energy state, the probability amplitudes for the production of an electron-positron pair with a total spin projection of $S_z=0$ and $S_z=1$ can be described by $c_{2,2}$ and $c_{1,2}$, respectively.

Correspondingly, the single-particle momentum distribution function of created electrons, $f_{s,s'}(\mathbf{q},t)$, denotes the probability for producing an electron with spin $s$ and a positron with spin $\overline{s'}$.
Finally, we can also obtain the momentum distribution $f_1(\mathbf{q}, t) =f_{1,1}(\mathbf{q}, t)+f_{1,2}(\mathbf{q}, t)$ for produced spin-up electrons, $f_2(\mathbf{q}, t)=f_{2,1}(\mathbf{q}, t)+f_{2,2}(\mathbf{q}, t)$ for produced spin-down electrons, and the total momentum distribution $f(\mathbf{q}, t)=f_1(\mathbf{q}, t)+f_2(\mathbf{q}, t)$.

\section{NUMERICAL RESULTS}
\label{sec:three}
The external field we considered is a spatially homogeneous and time-dependent elliptically polarized electric field,
\begin{equation}\label{eq:EPSEF}
\mathbf{E}\left( t \right) =E_0\sin^4\left(\frac{\Omega t}{2\tau}\right)\left[ \cos \left( \Omega t \right) \mathbf{e}_x+\delta \sin \left( \Omega t \right) \mathbf{e}_y \right],
\end{equation}
where $E_0$ is the amplitude of the electric field, $\Omega$ is the frequency, $\tau$ denotes the number of cycles, and $\mathbf{e}_x$ and $\mathbf{e}_y$ are the unit vectors in the $x$ and $y$ directions, respectively.
The elliptical polarization degree $\delta$ has a range of $[-1, 1]$.
For $\delta=+1$ and $\delta=-1$, the corresponding $z$-components of the photon spin angular momentum are $+1$ and $-1$, respectively.
For $\delta=0$, Eq. (\ref{eq:EPSEF}) corresponds to a linearly polarized electric field, and the average angular momentum of the photon is $0$.
Note that to ensure the convergence and validity of the calculation results, the electric field $\mathbf{E}(t_\mathrm{out})$ should be asymptotically approaching $0$ at final time, and in particular, its magnitude should be significantly smaller than that of the momentum distribution function.

\subsection{Results for boson pair production}\label{subsec:C}
\begin{figure}[!ht]
\centering
\includegraphics[width=0.99\columnwidth]{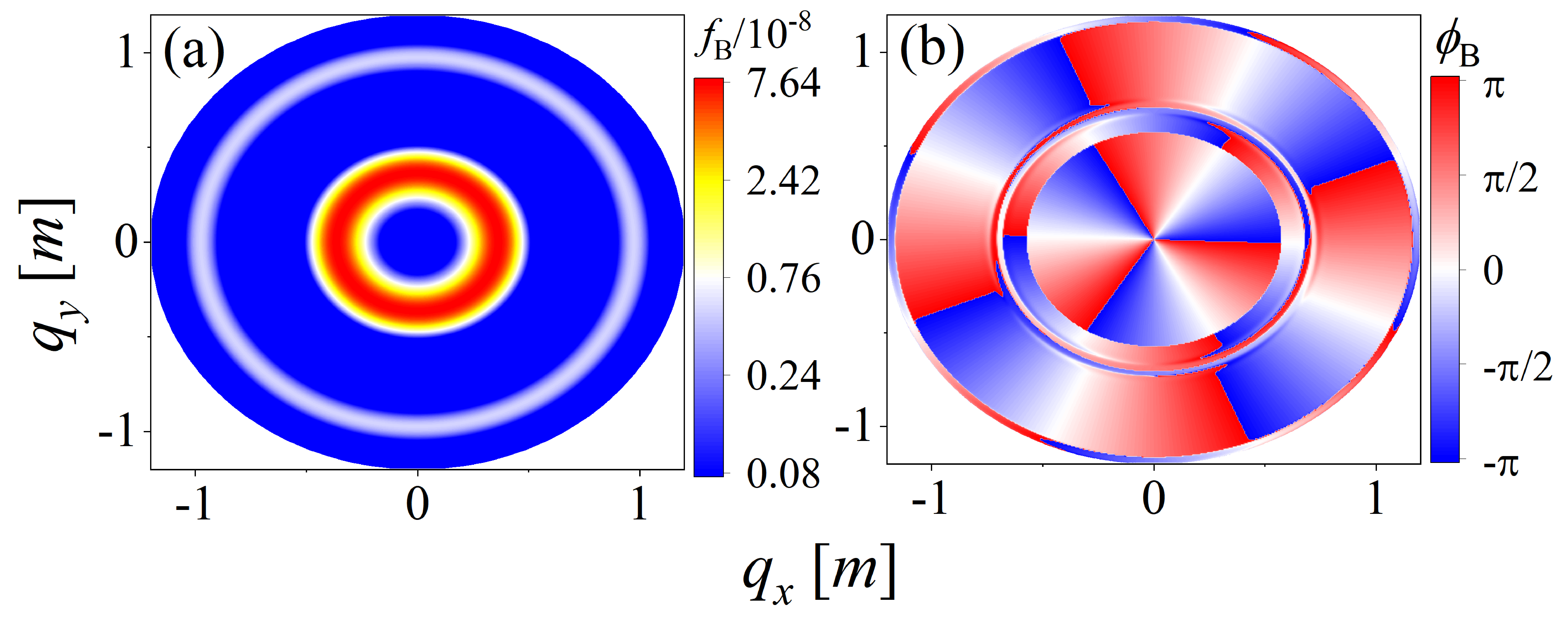}%
\par\medskip
\includegraphics[width=0.99\columnwidth]{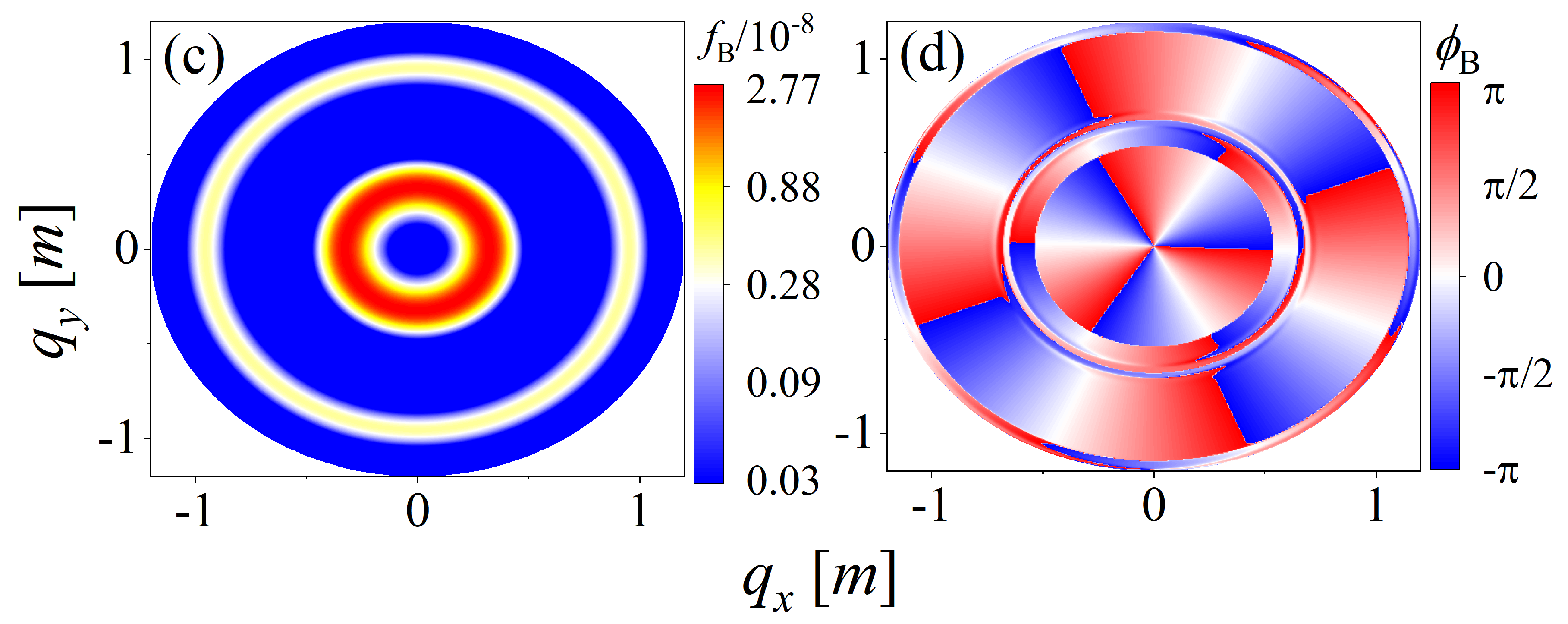}%
\caption{The momentum distributions (first column) and the phase distributions of the probability amplitude (second column) for produced bosons.
The first and second rows correspond to $q_z=0$ and $q_z=0.2m$, respectively.
The electric field parameters are $E_0=0.05E_{\mathrm{cr}}$, $\Omega=0.7m$, $\tau=20$, and $\delta=+1$.
\label{fig:fbargb}}
\end{figure}

The single-particle momentum distribution function $f_{\mathrm{B}}(\mathbf{q}, t_{\mathrm{out}})$ and the phase distribution of the probability amplitude $\phi _{\mathrm{B}}(\mathbf{q}, t_{\mathrm{out}})=\mathrm{arg}[c_{\mathrm{B}}(\mathbf{q}, t_{\mathrm{out}})]$ for bosons produced by the electric field Eq. (\ref{eq:EPSEF}) are shown in the first and second columns of Fig. \ref{fig:fbargb}, respectively.
The first and second rows of Fig. \ref{fig:fbargb} correspond to $q_z=0$ and $q_z=0.2m$, respectively.
Comparing Figs. \ref{fig:fbargb} (a) and (b) in this study with Figs. 2(c) and (f) in Ref. \cite{Fan2024nsl}, one can see that our results are consistent with those obtained using the two-level method in Ref. \cite{Fan2024nsl}, which confirms the reliability of our computational approach.
The circular ring structure exhibited by the momentum distribution in Figs. \ref{fig:fbargb}(a) and (c) is characteristic of the multiphoton absorption process.
For $N$-photon absorption process, we can approximately determine the radius of the multiphoton ring by the formula $q^N=[(N\Omega/2)^2-m_{\ast}^{2}]^{1/2}$ obtained from the energy conservation equation of pair production, where $m_{\ast}\approx m\sqrt{1+1/\gamma_{\mathrm{K}}^{2}}$ represents the effective mass of the particle \cite{Kohlfurst2013E}.
For instance, in Fig. \ref{fig:fbargb}(a), the radii of the multiphoton rings are approximately $0.664m$ and $1.238m$, corresponding to $3$-photon and $4$-photon absorption, respectively.

To understand the results in Figs. \ref{fig:fbargb}(b) and (d), we briefly review the concept of vortices.
In three-dimensional momentum space, vortices exhibit as continuous curves or closed loops, also known as vortex lines \cite{Bliokh2007ec,Bliokh2011fi,Bliokh2012az,Majczak2022xlv}.
Although the module of the probability amplitude vanishes along this line, the phase of the amplitude around it shifts by $2\mathrm{\pi}l$, where the integer $l=0, \pm 1, \pm 2, \dots$ is referred to as the topological charge \cite{Lloyd2017ipi,Bliokh2017uvr}.
In comparison, nodes manifest as surfaces with $0$ probability amplitude modulus, but traversing this surface results in a phase shift of $\pm \pi$.
The distinction between vortex lines and nodal surfaces was studied in detail in Ref. \cite{Geng2020V}.
Similar to Refs. \cite{Majczak2022xlv,Geng2020V}, vortex characteristics can be quantized using the following quantization condition
\begin{eqnarray}\label{eq:QVCBQC}
\oint_{\mathcal{C}}{\mathcal{A}}\left( \mathbf{q} \right) \cdot d\mathbf{q}=2\pi l,
\end{eqnarray}
where $\mathcal{C}$ is any closed non-contractible contour encircling the target vortex line in momentum space, $\mathcal{A}$ is the Berry connection, defined as $\mathcal{A} \left( \mathbf{q} \right) =\frac{1}{\left| c_{\mathrm{P}} \right|^2}\mathrm{Re}\left[ c_{\mathrm{P}}^{*}\left( -i\nabla _{\mathbf{q}} \right) c_{\mathrm{P}} \right]$ (or equivalently, the phase gradient of the probability amplitude $c_\mathrm{P}$ in momentum space).
As shown in Figs. \ref{fig:fbargb}(b) and (d), the phase distributions can provide information about the topological charge of the produced particles.
The phase shift on the closed path of the $3$-photon (or $4$-photon) absorption ring is $6\pi$ (or $8\pi$), so the topological charge is $3$ (or $4$).
Generally, for the $N$-photon absorption process, the topological charge of bosons is either $+N$ (when the electric field polarization is $\delta=+1$) or $-N$ (when the electric field polarization is $\delta=-1$).


Recently, Fan et al. \cite{Fan2024nsl} found that the topological charge read out from the phase distribution of the probability amplitude for boson pair production equals the intrinsic orbital angular momentum (IOAM) of produced bosons.
The IOAM refers to the OAM carried by the relative motion of the constituents of a composite system around their center of mass, as observed in the rest frame of that center of mass.
Since the momentum of a particle-antiparticle pair produced in a spatially uniform electric field is zero, the IOAM is equivalent to the OAM of the pair.
Furthermore, it is important to emphasize that the particle and antiparticle created from the vacuum are correlated rather than independent.
Consequently, the momentum distribution and the phase distribution of the probability amplitude we calculated for the particle or antiparticle are, in essence, projections of the global properties of this two-body system onto single-particle variables.
Therefore, the topological charge extracted from the phase distribution of the probability amplitude ($c_\mathrm{B}$ and $c_{s,s'}$) for the particle corresponds precisely to the $z$-component of the OAM of the particle-antiparticle pair.
This indicates that the topological charge reflects the OAM carried by the produced particle pairs and provides a direct means for quantifying and examining angular momentum transfer in vacuum pair production.

\subsection{Results for electron-positron pair production}\label{subsec:D}
\begin{figure}[!ht]
\centering
\includegraphics[width=0.99\columnwidth]{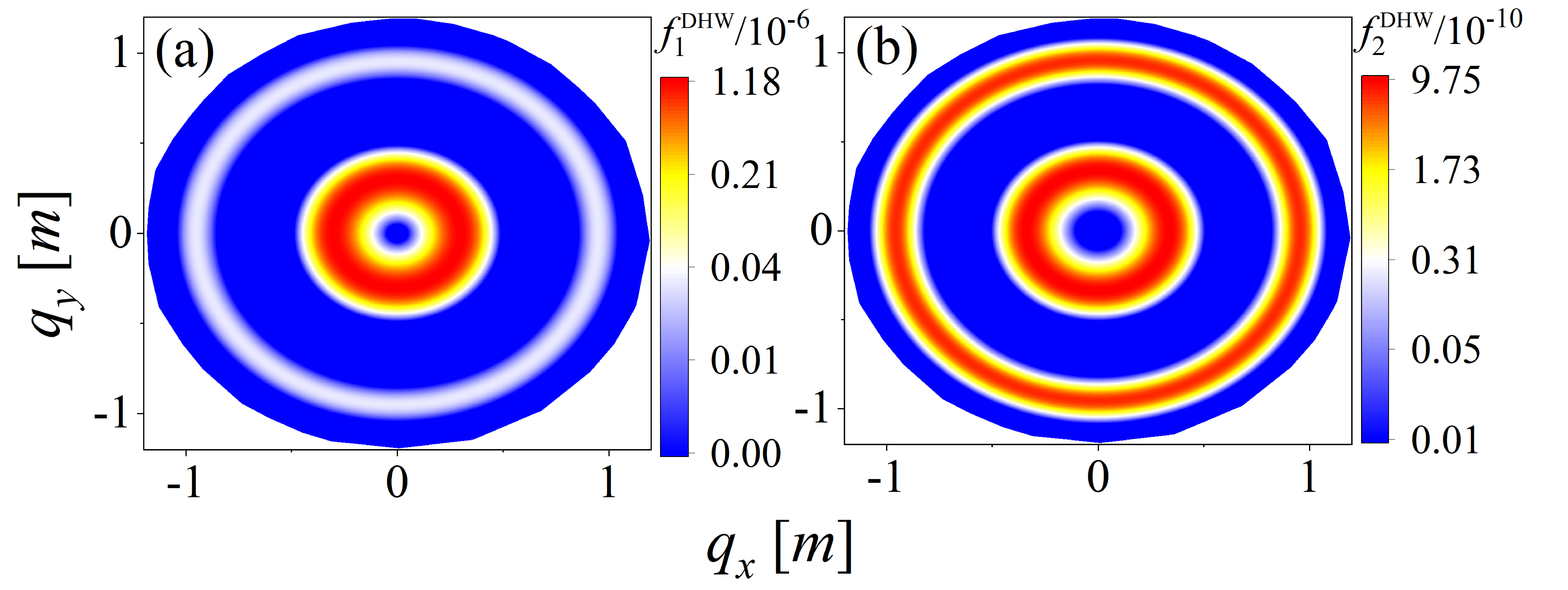}%
\par\medskip
\includegraphics[width=0.99\columnwidth]{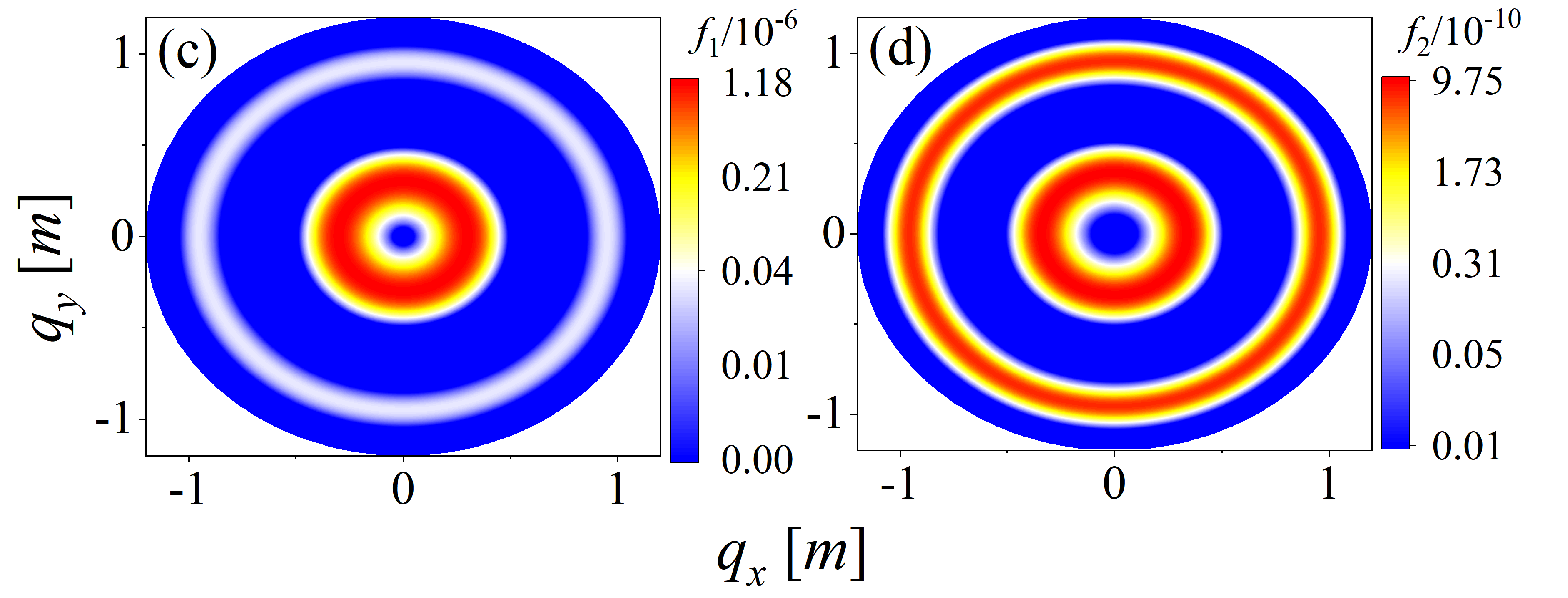}%
\caption{A comparison between the computational results from our method (second row) and those obtained using the DHW formalism (first row).
The left and right columns correspond to spin-up and spin-down particles, respectively.
The momentum $q_z$ equals $0.2m$.
The electric field parameters are $E_0=0.05E_{\mathrm{cr}}$, $\Omega=0.7m$, $\tau=20$, and $\delta=+1$.}
\label{fig:fDHWCQFT}
\end{figure}

To verify the reliability of our approach for studying electron-positron pair production, we show a comparison between the computational results obtained using our method and those obtained using the DHW formalism, see Fig. \ref{fig:fDHWCQFT}.
Here, $f_{1}^{\mathrm{DHW}}$ and $f_{2}^{\mathrm{DHW}}$ represent the momentum distribution functions for created spin-up and spin-down particles, respectively, within the DHW formalism framework.
The calculation method for $f_{1}^{\mathrm{DHW}}$ and $f_{2}^{\mathrm{DHW}}$ can be found in Appendix \ref{appa}.
As shown in Fig. \ref{fig:fDHWCQFT}, the results obtained by these two methods are identical.
Therefore, compared to the two-level model \cite{Amat2024nvg}, the method introduced in this paper provides a more accurate calculation of the spin-resolved momentum distribution function.

\begin{figure*}[t]
\centering
\includegraphics[width=0.98\textwidth]{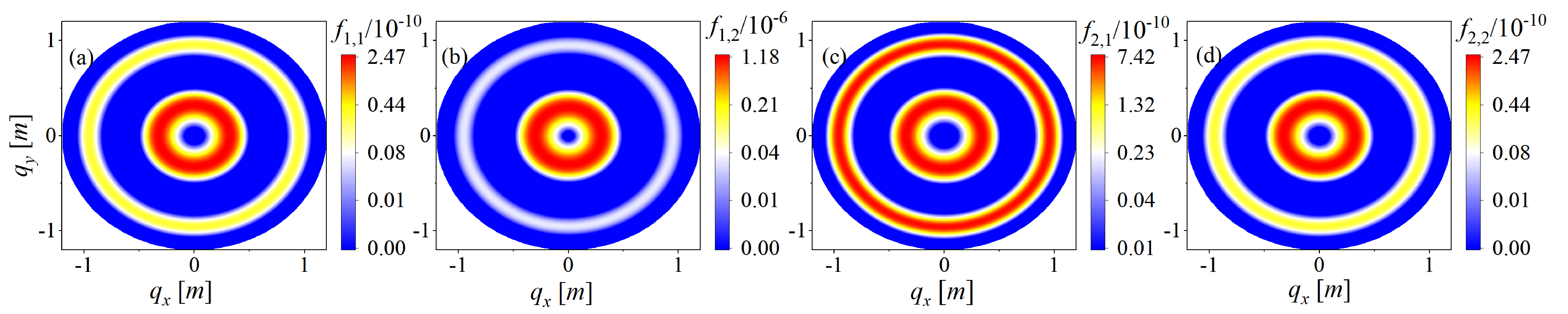}
\includegraphics[width=0.98\textwidth]{fFargpz02.eps}
\caption{The four spin-resolved momentum distribution functions $f_{s,s^{\prime}}$ (first row) and the phase distributions of the probability amplitude $\phi _{s,s^{\prime}}$ (second row) for produced electron-positron pairs.
Here, $s$ (or $s^{\prime}$) equals $1$ and $2$, representing spin up and spin down, respectively.
The momentum $q_z$ is $0.2m$.
The electric field parameters are $E_0=0.05E_{\mathrm{cr}}$, $\Omega=0.7m$, $\tau=20$, and $\delta=+1$.
\label{fig:fffarg}}
\end{figure*}

In Ref. \cite{Fan2024nsl}, the authors demonstrated that angular momentum is conserved in multiphoton boson pair production by analyzing the topological charge in the phase distribution of the probability amplitude for boson pair production.
Specifically, the spin angular momentum of the absorbed photons is entirely transferred to the OAM of the produced boson pairs.
However, two questions arise here.
First, the angular momentum conservation discussed in Ref. \cite{Fan2024nsl} is based on calculations of the $z$-components of the angular momentum of the field and the particle pairs.
Can it be extended to total angular momentum conservation?
Second, some may argue that the computed results presented in that work do not even demonstrate the conservation of the $z$-component of angular momentum, let alone total angular momentum conservation, because the calculated probability amplitude is for pair production, rather than for a particle or antiparticle individually.
The topological charge may not be related to the OAM of the particle or antiparticle but merely reflects that each absorbed photon contributes a phase factor to the probability amplitude for pair production.
To clarify angular momentum conservation in vacuum pair production and to investigate the transfer of angular momentum between the external field and the produced particles, we calculate the momentum distributions $f_{s,s^{\prime}}(\mathbf{q}, t_{\mathrm{out}})$ and the phase distributions of the probability amplitude $\phi_{s,s^{\prime}}=\mathrm{arg}[c_{s,s^{\prime}}(\mathbf{q}, t_{\mathrm{out}})]$ for fermion pair production in four spin configurations (see Fig. \ref{fig:fouramp}), as presented in Fig. \ref{fig:fffarg}.


Similar to the analysis of bosons, we can determine that the circular rings from inside to outside in the momentum distribution functions of Figs. \ref{fig:fffarg}(a)-(d) correspond to $3$-photon absorption and $4$-photon absorption, respectively.
Note that this is a rough estimate based on the energy conservation formula.
A more precise analysis would reveal that, in panel (c), the radii of the $3$-photon and $4$-photon absorption rings are larger than those in panels (a) and (d), which in turn are larger than those in panel (b).
This is because the angular momentum of the produced particles also affects the threshold for pair production.
From Figs. \ref{fig:fffarg}(e)-(h), one can see that when electrons and positrons have different spin combinations, the phase shift along a closed path around the multiphoton absorption ring also differs.
For instance, for the $3$-photon absorption process, the changes in $\phi_{1,1}$, $\phi_{1,2}$, $\phi_{2,1}$, and $\phi_{2,2}$ are $6\pi$, $4\pi$, $8\pi$, and $6\pi$, respectively, resulting in topological charges of $3$, $2$, $4$, and $3$.
This result indicates that multiphoton pair production in a circularly polarized field necessarily involves the angular momentum of the field and the particles.
If each absorbed photon merely contributed a phase factor to the probability amplitude for pair production, then all four cases discussed above would correspond to a topological charge of $3$, and topological charges of $2$ and $4$ would not appear.
The reason topological charges of $2$ and $4$ emerge is that the $z$-components of the spin angular momentum of the produced electron-positron pairs are $+1$ and $-1$, respectively.
This indicates that the spin angular momentum of the electron-positron pair also contributes a phase factor to the probability amplitude for pair production---or, in other words, the topological charge of the probability amplitude for pair production is precisely a manifestation of the angular momentum of the pair.
Moreover, for a circularly polarized field rotating counterclockwise about the $z$-axis, the $z$-component of the spin angular momentum of each photon is $+1$.
Hence, in the $3$-photon absorption process, the $z$-component of the total spin angular momentum acquired from the photons is $+3$.
Given that the $z$-components of the spin angular momentum of the electron-positron pairs are $0$, $+1$, $-1$, and $0$, respectively, if the $z$-component of the angular momentum is conserved in multiphoton pair production, the $z$-components of the OAM acquired by the electron-positron pairs should be $3$, $2$, $4$, and $3$, respectively.
These values are exactly the topological charges obtained from the probability amplitude for pair production.
Therefore, the topological charge in the probability amplitude for pair production is indeed a manifestation of the OAM of the produced electron-positron pairs.
Conversely, if we assume that the topological charge manifests the OAM of the electron-positron pairs, then we can demonstrate that the $z$-component of the angular momentum is conserved in multiphoton pair production process.
These two assumptions and their corresponding analyses are mutually corroborative.
Thus, we ultimately conclude that the topological charge in the probability amplitude for pair production reflects the OAM of the electron-positron pairs, and that the $z$-component of the angular momentum is conserved in multiphoton pair production.

The above discussion on angular momentum conservation was carried out separately for the four spin configurations.
In the following, we demonstrate the conservation of the $ z$-component of angular momentum during the pair production process by calculating the density of the $z$-component of the total angular momentum of produced pairs throughout the $N$-photon absorption process.
We denote the momentum distribution function, the $z$-component of the spin angular momentum, and the $z$-component of the OAM of the particle pairs produced via the $N$-photon absorption in the four spin configurations as $f^N_{s,s'}$, $S^N_{s,s'}$, and $l^N_{s,s'}$, respectively.
Then the density of the $z$-component of the total spin angular momentum of the produced electron-positron pairs is
\begin{equation}\label{eq:Sz}
S_z^{\mathrm{(e^+e^-)}} =\!\sum_{s,s^{\prime}}S^N_{s,s'}\!\int\!\left[ dq \right]f^N_{s,s'}=\!\int{\left[ dq \right] \left( f^N_{1,2}-f^N_{2,1} \right)}, \nonumber
\end{equation}
and that of the total OAM is
\begin{equation}\label{eq:Lz}
\begin{aligned}
L_z^{\mathrm{(e^+e^-)}}=&\!\sum_{s,s^{\prime}} l^N_{s,s^{\prime}}{\!\int\!\left[ dq \right]f^N_{s,s^{\prime}}} \\
=& \left( N-1 \right)\!\int\!\left[ dq \right] f^N_{1,2}+\left( N+1 \right) \int{\left[ dq \right] f^N_{2,1}}\\
&+ N\!\int\!\left[ dq \right] (f^N_{1,1}+f^N_{2,2}).\nonumber
\end{aligned}
\end{equation}
The density of the $z$-component of the total angular momentum is
\begin{equation}\label{eq:Jz}
\begin{aligned}
J_z^{\mathrm{(e^+e^-)}}=&S_z^{\mathrm{(e^+e^-)}}+L_z^{\mathrm{(e^+e^-)}}\\
=&N\!\int\!\left[ dq \right] (f^N_{1,1}+f^N_{1,2}+f^N_{2,1}+f^N_{2,2})\\
=&N\!\int\!\left[ dq \right] f^N.
\end{aligned}
\end{equation}
According to energy conservation in the $N$-photon absorption process, we have the energy conservation relation $N\Omega=2\omega_\mathbf{q}(t_\mathrm{out})$ and the number density of absorbed photons $n^N_{\gamma}=\frac{1}{\Omega}\int [dq]2\omega_\mathbf{q}(t_\mathrm{out})f^N=N\int [dq]f^N$.
Since the $z$-component of the spin angular momentum of each photon is $+1$, the density of the $z$-component of the total spin angular momentum of absorbed photons is
\begin{equation}\label{eq:Szgamma}
S_z^{\mathrm{(\gamma)}}=N\!\int\!\left[ dq \right] f^N.
\end{equation}
From Eqs. (\ref{eq:Jz}) and (\ref{eq:Szgamma}), we can obtain $S_z^{\mathrm{(\gamma)}}=S_z^{\mathrm{(e^+e^-)}}+L_z^{\mathrm{(e^+e^-)}}$, which indicates that the $z$-component of the angular momentum is conserved in multiphoton pair production.

This conservation can also be proved from the perspective of symmetry.
For the electric field (\ref{eq:EPSEF}), it is easy to show that the composite operator  $\widehat{C}=\widehat{J}_z-\widehat{H}(t)/\Omega$ is a conserved quantity, because it is commutative with the Hamiltonian $\widehat{H}(t)$.
Since there are no particle pairs in the initial state, we have $C=0$.
For an $N$-photon absorption process, energy conservation requires that the total energy $ 2\omega_{\mathbf{q}}(t_{\text{out}})$ of the produced pair equals the energy $N\Omega$ of the absorbed photons.
Consequently, for the final state, the $z$-component of the total angular momentum of the produced pair is $J_z = H(t_{\text{out}})/\Omega = 2\omega_{\mathbf{q}}(t_{\text{out}})/\Omega = N$.
Given that the $z$-component of the spin angular momentum of each photon is $+1$, we can conclude that the $z$-component of the total angular momentum of the field and the particle pair is conserved in multiphoton pair production.
However, since there are no analogous conserved composite operators along the $x$ and $y$ directions, the conservation of total angular momentum cannot be verified.
Note these results are obtained within the semiclassical framework we used, where the electromagnetic field is treated as a classical background.
In a fully quantized framework, angular momentum remains conserved during multiphoton pair production, because the fully quantized QED Lagrangian possesses global spatial rotation invariance.

In addition, Figs. \ref{fig:fffarg}(a)-(d) also shows that the maxima of the four spin-resolved momentum distribution functions $f_{s,s'}^\mathrm{max}$ satisfy the relation $f_{1,2}^\mathrm{max}>f_{2,1}^\mathrm{max}>f_{1,1}^\mathrm{max}
=f_{2,2}^\mathrm{max}$.
The corresponding particle number density also satisfies this relation.
It is easy to understand that the number density of produced particle pairs with the $z$-component of the spin angular momentum $S_z=+1$ is the largest.
Because, in general, the smaller the total OAM of the produced particle pairs, the lower the energy required for their production, and thus the more easily they are produced.
However, the number density of pairs produced with the $z$-component of the spin angular momentum $S_z=0$ is smaller than that for $S_z=-1$, although the OAM for the former is smaller than that for the latter.
In particular, when $q_z = 0$, the number of produced pairs with $S_z=0$ is zero.
As $q_z$ increases, the fraction of such pairs in the total number density also increases.
This indicates that, besides the influence of the OAM of the particles, other factors also affect the pair production process.
The factor is the conservation of charge-conjugation parity in pair production.
The two cases with $S_z=0$ correspond to the direct product states of an electron-positron pair with antiparallel spins, which constitute the $S_z=0$ state of the spin triplet rather than the spin singlet, because electron-positron pairs produced in a spatially uniform oscillating electric field are in spin-symmetric states ($S=1$) \cite{Mocken2010}.
The $C$-parity of each photon is $-1$, and the $C$-parity of an electron-positron pair with OAM quantum number $L$ is $(-1)^{L+S}=(-1)^{L+1}$.
Therefore, $C$-parity conservation in an $N$-photon absorption process requires $(-1)^N = (-1)^{L+1}$.
In the $3$-photon absorption process, for $S_z=0$, we have $N=3$ and $L=3$, which does not satisfy $C$-parity conservation.
Thus, this process is suppressed.
Especially, when $q_z = 0$, the pair production process is dominated by the production of pairs with OAM quantum number $L=3$, which is strictly forbidden by $C$-parity conservation.
When $q_z \neq 0$, pairs with larger $L$ also begin to be produced, so the production process is not completely forbidden but rather suppressed.
As $q_z$ increases, this suppression weakens, leading to an increase in the number density fraction of produced pairs with $S_z=0$.

Moreover, we can also explain from the perspective of spin-field coupling why the number of produced electron-positron pairs with $S_z=0$ is smaller than that with $S_z=-1$.
It is known that the electric field (\ref{eq:EPSEF}) in the laboratory frame can generate an effective magnetic field $\mathbf{B}'(t) = -\gamma_v \mathbf{v} \times \mathbf{E}(t)$ in the rest frame of a particle, where $\gamma_v$ is the Lorentz factor.
The coupling between spin and this magnetic field is described by the Hamiltonian $H_\mathrm{int}= -\boldsymbol{\mu} \cdot \mathbf{B}'$, where $\boldsymbol{\mu}\approx\mp\frac{e}{m}\mathbf{S}$ are magnetic moments of electrons and positrons, respectively.
For an electron, the coupling energy is lowest when its spin is antiparallel to the effective magnetic field, making production most favorable in this configuration.
For a positron, the coupling energy is lowest when its spin is parallel to the effective magnetic field.
However, since the electron and positron have opposite velocities, the effective magnetic fields they experience are also opposite. Consequently, their spins tend to be aligned in the same direction,  that is, pair production favors a parallel spin configuration.
When $q_z = 0$, the magnetic field is always along the $z$-axis and varies with time, leading to electron and positron spins aligned parallel along the $z$-axis.
Hence, the number of pairs with $S_z= 0$ is zero.
When $q_z \neq 0$, the magnetic field rotates about the $z$-axis. The presence of a magnetic field component perpendicular to the $z$-axis allows the spins to deviate from the $z$-axis, thereby permitting the appearance of antiparallel spin configurations.
For the multiphoton absorption process, a larger $q_z$ corresponds to smaller $p_x$ and $p_y$.
This results in a smaller component of the magnetic field along the $z$-axis and a larger component perpendicular to it.
Therefore, the number density fraction of produced pairs with $S_z = 0$ increases.

In contrast to electron-positron pairs, which are produced in spin-symmetric states, boson pairs have a spin angular momentum quantum number $S=0$.
Consequently, their production process is not suppressed by $C$-parity conservation, see Fig. \ref{fig:fbargb}(c).
Comparing boson pair production with the production of electron-positron pairs having the $z$-component of the spin angular momentum $S_z = \pm 1$ displayed in Fig. \ref{fig:fffarg}, we find that the number of produced boson pairs is significantly larger than that of electron-positron pairs with $S_z = -1$, but considerably smaller than that of pairs with $S_z = 1$.
This is because the OAM of the produced boson pairs is smaller than that of electron-positron pairs with $S_z=-1$ but larger than that of pairs with $S_z=1$.
The above result shows that the larger the OAM of the produced particle pairs, the less likely the pair production process is to occur.
That is, pair production favors the production of particle pairs with smaller OAM.

\section{SUMMARY}
\label{sec:four}

In this study, we focus on a method for accurately calculating the momentum distribution and the phase distribution of the probability amplitude for pair production, as well as the angular momentum transfer between external fields and produced particles.

We theoretically derive the probability amplitudes for boson and spin-resolved fermion pair production from the Klein-Gordon equation and the Dirac equation, respectively.
We demonstrate that our method ensures that the phase distribution of the probability amplitude is well resolved without guessing the dynamic phase factor.
By analyzing the probability amplitudes for spin-resolved fermion pair production within the framework of the Dirac sea, we find that they correspond to four distinct spin configurations of the electron and positron: both up, both down, and two cases with opposite spins.

The angular momentum transfer in multiphoton pair production under a spatially homogeneous and time-dependent circularly polarized electric field is investigated numerically.
In this field configuration, each absorbed photon contributes $\pm 1$ unit of angular momentum along the $z$-axis.
From the phase distribution of the probability amplitude for the  particle or the antiparticle, we extract the topological charge---a vortex number reflecting the OAM of the produced pairs.
We clarify that this topological charge represents the OAM of the entire particle-antiparticle pair rather than that of an individual particle.
Based on this insight, we demonstrate that, for all four spin combinations of electron-positron pairs, the $z$-component of the total angular momentum of the field and the particles is conserved, whereas the total angular momentum conservation cannot be verified.
Moreover, the results show that the pair production process is also constrained by $C$-parity conservation, and that pairs with smaller OAM are produced more favorably.
These findings indicate that the OAM of the pair can be inferred by measuring the topological charge of either the particle or its antiparticle, offering a new perspective on angular momentum transfer from the field to produced particles in the pair production process.

\begin{acknowledgments}
We are grateful to H. H. Fan and B. S. Xie for helpful discussions. The work is supported by the National Natural Science Foundation of China (NSFC) under Grants No. 11974419 and No. 11705278, the Strategic Priority Research Program of Chinese Academy of Sciences (Grant No. XDA25051000, XDA25010100), and the Fundamental Research Funds for the Central Universities (No. 2023ZKPYL02, 2025 Basic Sciences Initiative in Mathematics and Physics).
\end{acknowledgments}

\appendix

\section{Dirac-Heisenberg-Wigner (DHW) formalism}\label{appa}
In this appendix, we briefly introduce the method for calculating the spin-resolved momentum distribution function in a framework of the DHW formalism under spatially homogeneous conditions.
The single-particle momentum distribution function $f^{\mathrm{DHW}}(\mathbf{q},t)$ can be obtained by solving the following system of ordinary differential equations \cite{Li2017M,Blinne2015zpa}:
\begin{equation}\label{eq:DHWEQS}
\begin{split}
&\dot{f}^{\mathrm{DHW}}=\frac{e\mathbf{E}\cdot\mathbf{v}}{2\omega},
\\
&\dot{\mathbf{v}}=\frac{2\left[ \left( e\mathbf{E}\cdot \mathbf{p} \right) \mathbf{p}-\omega ^2e\mathbf{E} \right]}{\omega ^3}\left( f^{\mathrm{DHW}}-1 \right)
\\
&\quad \quad -\frac{\left( e\mathbf{E}\cdot \mathbf{v} \right) \mathbf{p}}{\omega ^2}-2\mathbf{p}\times \mathbf{a}-2m\mathbf{t}_1,
\\
&\dot{\mathbf{a}}=-2\mathbf{p}\times \mathbf{v},
\\
&\dot{\mathbf{t}}_1=\frac{2}{m}\left[ m^2\mathbf{v}+\left( \mathbf{p}\cdot \mathbf{v} \right) \mathbf{p} \right],
\end{split}
\end{equation}
with the initial conditions $f^{\mathrm{DHW}}\left( \mathbf{q},-\infty \right) =0$, and $\mathbf{a}\left( \mathbf{q},-\infty \right) =\mathbf{v}\left( \mathbf{q},-\infty \right) =\mathbf{t}_1\left( \mathbf{q},-\infty \right)=0$.

According to Ref. \cite{Jiang2025kmq}, the spin-resolved momentum distribution function is
\begin{align}\label{eq:SRMDODHW}
f_{1,2}^{\mathrm{DHW}}\!=\!\frac{f^{\mathrm{DHW}}}{2}\!\mp\! \frac{\boldsymbol{\eta }}{4\omega}\cdot \left[ m\mathbf{a}\!+\!\mathbf{p}\!\times\! \mathbf{t}_1\!+\!\frac{\left( \mathbf{p}\cdot \mathbf{a} \right) \mathbf{p}}{\omega +m} \right],
\end{align}
where $\boldsymbol{\eta}$ is the spin-direction unit vector in the rest frame of the particle.
In this paper, we study the case where $\boldsymbol{\eta}=(0,0,1)$.

\end{document}